\documentstyle[12pt,epsf]{article}
\newcommand{\ul}[1]{\underline{#1}}
\newcommand{\sla}[1]{/\!\!\!\!{#1}}
\newcommand{\I}{{\rm I}}
\newcommand{\II}{{\rm I\!I}}
\newcommand{\bein}[1]{e^{\underline{#1}}}
\newcommand{\ten}{\natural}

        \newcommand{\comment}[1]{}

\begin{document}
\begin{flushright}
UT-873\\ 
February, 2000
\end{flushright}
\vskip2cm
\begin{center}
\vskip1cm
\noindent{\Large \bf On Branes Ending on Branes in Supergravity}
\vskip4mm
\noindent{\it \large Kazuo Hosomichi\footnote
{\tt email address: hosomiti@hep-th.phys.s.u-tokyo.ac.jp}}\\
\vskip3mm
{\small\it Department of Physics, Faculty of Science, University of Tokyo,\\
     Hongo 7-3-1, Bunkyo-ku, Tokyo 113-0033, Japan}
\vskip4mm
\end{center}
\vskip1cm
\begin{abstract}
 We study the eleven-dimensional supergravity
 and the classical solutions corresponding to
 M2-branes ending on M5-branes.
 We obtain the generic BPS configuration
 representing two sets of parallel M5-branes
 with $(1\!+\!1)$ commonly longitudinal directions
 and M2-branes stretching between them.
 We also discuss how the brane creation
 is described in supergravity.
\end{abstract}

\newpage

\section{Introduction}
   {\it Branes}
 or the spatially extended solitons
 play the central role in the study
 of string theory or M-theory.
 They are known to form a variety of bound states,
 and the corresponding classical solutions should exist
 in the effective supergravity theories.
 However, a difficulty arises in constructing
 classical solutions corresponding to the bound states
 of intersecting branes.
 The solutions for two intersecting branes have been found
 only when one of the two branes are smeared
 along the directions longitudinal to the
 other\cite{Gauntlett:1996pb,Tseytlin:1997cs}.
 (There are a few exceptions such as
  \cite{Khuri:1993ii} and \cite{Hashimoto:1999ug})
 Recently a fully localized solutions have been found
 for M5-branes intersecting on three-branes\cite{Fayyazuddin:1999zu}
 and for M2-brane junctions\cite{Ramadevi:1999dy}.
 There has also been a perturbative analysis
 of Born-Infeld theory coupled to
 supergravity\cite{Gomberoff:2000ps}.
 In this paper we try to obtain
 another solutions for localized branes
 intersecting one another.

   In this paper we shall work
 in the eleven-dimensional supergravity\cite{Cremmer:1978km}.
 It contains a three-form potential
 besides the graviton and the gravitino,
 and M2-branes and M5-branes are
 the electric and magnetic sources of the potential.
 The presence of a single BPS M2-brane or M5-brane
 preserves a half of supersymmetry.
 An M2-brane and an M5-brane are known to form
 a $1/4$-supersymmetric bound state
 when they intersect on a string.
 It is also believed that an M2-brane can end on an M5-brane.
 In this case the M2-brane is ``open''
 and has a $(1\!+\!1)$-dimensional boundary on the M5-brane.
 Interestingly, such open M2-branes can be created
 between two M5-branes with
 $(1\!+\!1)$ commonly longitudinal directions
 when the two M5-branes pass through each other.
 Such creation of branes are familiar
 in the D-brane worldvolume gauge field theory,
 and a supergravity analysis has been
 made for a specific example\cite{Gomberoff:1999zy}.

   In the following we shall find
 the generic BPS configuration
 corresponding to the bound states of
 M2-branes$(012)$, M5-branes$(013456)$
 and M5-branes$(01789\ten)$.
 The numbers specify the directions
 the branes are lying along,
 and we denoted the eleventh direction by $\ten$.
 The analysis of the BPS condition
 $\delta_{\rm SUSY} \psi_m\!=\!0$
 is performed in section 2
 for spherical symmetric configurations,
 and the result is fully generalized in section 3.
 The resultant expression depends on
 three arbitrary functions of $x_{2,3,\ldots,9,\ten}$.
 These functions have to satisfy
 the equations of motion given in section 4
 in order to describe a bound state of
 localized(delta-functional) branes.
 They are not solved, but we show that
 if the M5-branes are localized,
 the boundaries of M2-branes stretching between them
 are automatically localized.
 We also propose in section 5 an idea for understanding
 the brane creation in supergravity.

\section{Analysis of the BPS Condition}

   Here we analyze the BPS condition
 under the assumption of spherical symmetry.
 The solution is summarized in
 (\ref{spherical symmetric metric}) and
 (\ref{spherical symmetric field strength}),
 and the following arguments
 help the reader to understand
 to what extent the result is general.

   We parameterize
 01-directions   by $t,\sigma$,
 2-direction     by $x_2$,
 3456-directions by $x_1, \theta_\I, \phi_\I, \psi_\I$ and
 789$\ten$-directions by $x_3, \theta_\II, \phi_\II, \psi_\II$.
   We assume the following form for the metric:
\begin{equation}
  ds^2=g_0^2(-dt^2+d\sigma^2+
             \varpi^1\varpi^1+\varpi^2\varpi^2+\varpi^3\varpi^3)
      +g_\I^2d\Omega_\I^2+g_\II^2d\Omega_\II^2
\end{equation}
\[
  d\Omega_i^2= d\theta_i^2+\cos^2\theta_i d\phi_i^2+\sin^2\theta_i d\psi_i^2 
     ~~,~~~ i=\I ~{\rm or}~ \II 
\]
 where $\varpi^{1,2,3}$ are linear combinations of $dx^{1,2,3}$
 with coordinate-dependent coefficients.
 We define the vielbein as follows:
\[
\begin{array}{rclcrclclcrclcrcl}
  e^{\ul{t}}	        &=& g_0 dt			        &;&
  e^{\ul{1}}	        &=& g_0\varpi^1                         &;&
  e^{\ul{\theta_\I}}	&=& g_\I  d\theta_\I		        &;&
  e^{\ul{\theta_\II}}	&=& g_\II d\theta_\II		        \\
  e^{\ul{\sigma}}	&=& g_0 d\sigma			        &;&
  e^{\ul{2}}	        &=& g_0\varpi^2   	                &;&
  e^{\ul{\phi_\I}}	&=& g_\I \cos\theta_\I d\phi_\I	        &;&
  e^{\ul{\phi_\II}}     &=& g_\II\cos\theta_\II d\phi_\II       \\ & & & &
  e^{\ul{3}}	        &=& g_0\varpi^3 	                &;&
  e^{\ul{\psi_\I}}	&=& g_\I \sin\theta_\I d\psi_\I	        &;&
  e^{\ul{\psi_\II}}     &=& g_\II \sin\theta_\II d\psi_\II
\end{array}
\]
 Hereafter we denote local Lorentz indices with underbar.
 Under the spherical symmetry, the most generic form for
 the four-form field strength is given by:
\begin{eqnarray}
  F_{(4)}&=&
  F_{(4)}^{[M2]}+F_{(4)}^{[M5]}+F_{(4)}^{[M5']} \nonumber \\ &=&
  g_0^{-1}\left(
  \frac{1}{2}\bein{t\sigma ij}E_k\epsilon^{ijk}
  +\bein{\theta_\II\phi_\II\psi_\II i}B_i
  +\bein{\theta_\I\phi_\I\psi_\I i}H_i
  \right) \nonumber \\
   &=&
   dtd\sigma\wedge\frac{1}{2}g_0^3\epsilon^{ijk}E_i \varpi^{jk}
  +d^3\Omega_\II\wedge g_\II^3 B_i \varpi^i
  +d^3\Omega_\I\wedge g_\I^3 H_i \varpi^i ~.
\label{Field Strength}
\end{eqnarray}
\comment{
 And its dual has the following components:
\begin{eqnarray*}
  F_{(7)}&=&
  F_{(7)}^{[M2]}+F_{(7)}^{[M5]}+F_{(7)}^{[M5']} \\ &=&
  g_0^{-1}\left(
  \bein{\theta_\I\phi_\I\psi_\I\theta_\II\phi_\II\psi_\II i}E_i
 -\frac{1}{2}\bein{t\sigma\theta_\I\phi_\I\psi_\I ij}B_k\epsilon^{ijk}
 +\frac{1}{2}\bein{t\sigma\theta_\II\phi_\II\psi_\II ij}H_k\epsilon^{ijk}
 \right) \\
  &=& 
   d^3\Omega_\I d^3\Omega_\II\wedge g_\I^3 g_\II^3 E_i\varpi^i
  -dtd\sigma d^3\Omega_\I\wedge
  \frac{1}{2}g_0^3g_\I^3\epsilon^{ijk}B_i\varpi^{jk} 
  +dtd\sigma d^3\Omega_\II\wedge
  \frac{1}{2}g_0^3g_\II^3\epsilon^{ijk}H_i\varpi^{jk} ~.
\end{eqnarray*}
 Here the Hodge dual is defined as follows:
\[
 \ast(\bein{m_1\cdots m_4})=
 \frac{1}{7!}\epsilon^{m_1\cdots m_{11}}e_{\ul{m_5\cdots m_{11}}}
 ~,~~~
 \epsilon^{\ul{t\sigma 123 \theta_\I \phi_\I \psi_\I
                           \theta_\II\phi_\II\psi_\II}} =1.
\]
 Using the above definition we can easily find
\begin{eqnarray*}
  \ast(\bein{t\sigma 12}) &=&
  \bein{\theta_\I\phi_\I\psi_\I\theta_\II\phi_\II\psi_\II 3} \\
  \ast(\bein{\theta_\II\phi_\II\psi_\II 1}) &=&
  -\bein{t\sigma \theta_\I\phi_\I\psi_\I 23} \\
  \ast(\bein{\theta_\I\phi_\I\psi_\I 1}) &=&
  \bein{t\sigma \theta_\II\phi_\II\psi_\II 23} ~.
\end{eqnarray*} 
 Be careful for the additional sign
 in raising or lowering the index $\ul{t}$.}
 Here the overall factor $g_0^{-1}$ in the second line
 is simply for later convenience.

   To obtain supersymmetric solutions
 we have to focus on the supersymmetry transformation law
 of gravitino
\begin{eqnarray}
  e_{\ul{m}}^{~m}\delta_\epsilon \psi_m &=&
  e_{\ul{m}}^{~m}{\cal D}_m\epsilon-\frac{1}{288}
  (3\Gamma^{\ul{pqrs}}\Gamma_{\ul{m}}-\Gamma_{\ul{m}}\Gamma^{\ul{pqrs}})
  \epsilon F_{\ul{pqrs}} ~,
\label{BPS condition}
\end{eqnarray}
 and find field configurations that admit
 $\delta_\epsilon\psi_m=0$ for some nonzero $\epsilon$.
 Here and throughout this section
 we use ``mostly Hermitian'' Gamma matrices satisfying
\[
   \left\{ \Gamma^{\ul{a}}, \Gamma^{\ul{b}} \right\}
   =2\eta^{\ul{ab}}=2{\rm diag}(-++\cdots +)~,~~
   \Gamma^{\ul{a_1\cdots a_{11}}}=\epsilon^{a_1\cdots a_{11}}~,~~
  \epsilon^{t\sigma
  123\theta_\I\phi_\I\psi_\I\theta_\II\phi_\II\psi_\II}=1.
\]

    Let us evaluate (\ref{BPS condition}) term by term.
  To begin with, the covariant derivative of a spinor is defined by
\begin{equation}
  e_{\ul{m}}^{~m}{\cal D}_m\epsilon =
  e_{\ul{m}}^{~m}(\partial_m
  +\frac{1}{4}\Omega_{m\ul{pq}}\Gamma^{\ul{pq}})\epsilon\equiv
  (\nabla_{\ul{m}}+\frac{1}{4}\Omega_{\ul{mpq}}\Gamma^{\ul{pq}})\epsilon
  ~~;~~~~
  \nabla_{\ul{m}}\equiv e_{\ul{m}}^{~m}\partial_m
\end{equation}
 where $\Omega_{\ul{pq}}\!=\!dx^m\Omega_{m\ul{pq}}$ is the spin connection.
 Under the assumption of spherical symmetry
 we can obtain some components of the spin connection
 from the torsion-free condition alone:
\begin{equation}
  {\cal D}e^{\ul{p}}= de^{\ul{p}}+\Omega^{\ul{p}}_{~\ul{q}}e^{\ul{q}}=0.
\label{torsion-free condition}
\end{equation}
 The only nonzero components of $\Omega_{\ul{mnp}}$
 except for those with $(m,n,p=1,2,3)$ are
\begin{eqnarray*}
 -\Omega_{\ul{tti}}=\Omega_{\ul{\sigma\sigma i}}
  &=&\nabla_{\ul{i}}\ln g_0 \\
  \Omega_{\ul{\theta_\I\theta_\I i}}=
  \Omega_{\ul{\phi_\I\phi_\I i}}=
  \Omega_{\ul{\psi_\I\psi_\I i}}
  &=&\nabla_{\ul{i}}\ln g_\I \\
  \Omega_{\ul{\theta_\II\theta_\II i}}=
  \Omega_{\ul{\phi_\II\phi_\II i}}=
  \Omega_{\ul{\psi_\II\psi_\II i}}
  &=&\nabla_{\ul{i}}\ln g_\II
\end{eqnarray*}
\begin{eqnarray*}
  \Omega_{\ul{\phi_\I\phi_\I\theta_\I}}=
  -\frac{1}{g_\I}\frac{\sin\theta_\I}{\cos\theta_\I} &;&
  \Omega_{\ul{\psi_\I\psi_\I\theta_\I}}=
  \frac{1}{g_\I}\frac{\cos\theta_\I}{\sin\theta_\I} \\
  \Omega_{\ul{\phi_\II\phi_\II\theta_\II}}=
  -\frac{1}{g_\II}\frac{\sin\theta_\II}{\cos\theta_\II} &;&
  \Omega_{\ul{\psi_\II\psi_\II\theta_\II}}=
  \frac{1}{g_\II}\frac{\cos\theta_\II}{\sin\theta_\II}
\end{eqnarray*}
 Using these expressions we can rewrite
 the BPS condition $\delta_\epsilon\psi_m\!=\!0$
 in the following way:
\begin{eqnarray}
  2\Gamma^{\ul{\theta_i}}\nabla_{\ul{\theta_i}}\epsilon&=&
  2\Gamma^{\ul{\phi_i}}\nabla_{\ul{\phi_i}}\epsilon
  +\Gamma^{\ul{\theta_i}}\epsilon\nabla_{\ul{\theta_i}}\ln\cos\theta_i 
  \nonumber \\ &=&
  2\Gamma^{\ul{\psi_i}}\nabla_{\ul{\psi_i}}\epsilon
  +\Gamma^{\ul{\theta_i}}\epsilon\nabla_{\ul{\theta_i}}\ln\sin\theta_i 
  ~~,~~ i=\I~{\rm or}~\II
\label{angular equation}
\end{eqnarray}
\begin{eqnarray}
  g_0(4\nabla_{\ul{i}}+\Omega_{\ul{ijk}}\gamma^{jk}
   -2\nabla_{\ul{j}}\ln g_0\gamma^i\gamma^j)\epsilon
 &=& 2(~~~E_i\gamma^{123}+B_i\gamma^0-H_i\gamma^5)\epsilon \nonumber \\
  3g_0\sla{\nabla}\ln g_0\epsilon &=&
  ~(-2\sla{E}\gamma^{123}
    -~~\sla{B}\gamma^{0}
    +~\sla{H}\gamma^{5})\epsilon \nonumber \\
  6g_0\Gamma^{\ul{t\sigma\theta_\I}}\nabla_{\ul{\theta_\I}}\epsilon+
  3g_0\sla{\nabla}\ln g_\I\epsilon &=&
  ~(~~~~\sla{E}\gamma^{123}
    -~~\sla{B}\gamma^{0}
    -2\sla{H}\gamma^{5})\epsilon \nonumber \\
  6g_0\Gamma^{\ul{t\sigma\theta_\II}}\nabla_{\ul{\theta_\II}}\epsilon+
  3g_0\sla{\nabla}\ln g_\II\epsilon &=&
  ~(~~~~\sla{E}\gamma^{123}
    +\,2\sla{B}\gamma^{0}
    +~\sla{H}\gamma^{5})\epsilon ~.
\label{x_123 equation}
\end{eqnarray}
 Here we have introduced a set of new gamma-matrices
\begin{equation}
  (\gamma^0,\gamma^1,\gamma^2,\gamma^3) =
  (\Gamma^{\ul{\theta_\II\phi_\II\psi_\II}},
   \Gamma^{\ul{t\sigma 1}},\Gamma^{\ul{t\sigma 2}},\Gamma^{\ul{t\sigma 3}})
  ~~,~~
  \gamma^5=\gamma^0\gamma^1\gamma^2\gamma^3=
  -\Gamma^{\ul{\theta_\I\phi_\I\psi_\I}}
\end{equation}
 and used some short-hand notations
\begin{equation}
  \sla{\nabla}\equiv
  \gamma^1\nabla_{\ul{1}}+\gamma^2\nabla_{\ul{2}}+\gamma^3\nabla_{\ul{3}} ~,~
  \sla{E}\equiv \gamma^1E_1+\gamma^2E_2+\gamma^3E_3 ~,~
  \rm{etc.}
\end{equation}
 $\gamma^0$ and $\gamma^5$ are anti-hermitian
 while $\gamma^{1,2,3}$ are hermitian.
 We also assumed in the above
 that $\epsilon$ does not depend on $t$ and $\sigma$.

~

 We would like to obtain the bosonic field configuration
 that satisfies the BPS condition $\delta_\epsilon\psi_m=0$
 with the following $\epsilon$:
\begin{equation}
 \epsilon=f(x_i)U_\I (\theta_\I, \phi_\I, \psi_\I)
                U_\II(\theta_\II,\phi_\II,\psi_\II)\epsilon_0,
\label{assumption for epsilon}
\end{equation}
 where $f(x_i)$ is a scale factor,
 $U_{\I,\II}$ are two mutually commuting
 local Lorentz transformations
 and $\epsilon_0$ is a constant spinor.
 We also assume that the residual supersymmetry is
 characterized by
\begin{equation}
 \Gamma^{\ul{t\sigma 2}}\epsilon=
 \Gamma^{\ul{t\sigma 3\theta_\II\phi_\II\psi_\II}}\epsilon=
 \Gamma^{\ul{t\sigma 1\theta_\I\phi_\I\psi_\I}}\epsilon=\epsilon
  ~~~~\mbox{or}~~~~
 \gamma^2\epsilon=\gamma^{30}\epsilon=\gamma^{51}\epsilon=\epsilon.
\label{residual SUSY}
\end{equation}
 We can rather easily find the solution
 of the angular equation (\ref{angular equation})
 satisfying also (\ref{residual SUSY}).
 It is given by the following $U_\I$ and $U_\II$:
\begin{eqnarray}
  U_{\I} &=&\exp(-\frac{\theta_\I\Gamma^{\ul{\theta_\I 1}}}{2})
            \exp(-\frac{\phi_\I  \Gamma^{\ul{\phi_\I 1}}}{2})
            \exp(-\frac{\psi_\I  \Gamma^{\ul{\psi_\I\theta_\I}}}{2})
  \nonumber \\
  U_{\II}&=&\exp(-\frac{\theta_\II\Gamma^{\ul{\theta_\II 3}}}{2})
            \exp(-\frac{\phi_\II  \Gamma^{\ul{\phi_\II 3}}}{2})
            \exp(-\frac{\psi_\II  \Gamma^{\ul{\psi_\II\theta_\II}}}{2})
\label{two local Lorentz}
\end{eqnarray}
 These local Lorentz transformations are understood as
 relating the polar frame
 (where $\Gamma^{\ul{t}, \ul{\sigma}, \ul{1}, \ldots,\ul{\psi_\II}}$
  are coordinate independent)
 to the orthonormal frame
 (where  $\Gamma^{\ul{0},\ul{1},\ldots,\ul{\ten}}$
  are coordinate independent).
\comment{
 The equation (\ref{angular equation}) can be rewritten as
 \begin{eqnarray*}
   2\cos\theta\Gamma^\theta\partial_\theta\epsilon &=&
   2\Gamma^\phi\partial_\phi\epsilon -\sin\theta\Gamma^\theta\epsilon \\
   2\sin\theta\Gamma^\theta\partial_\theta\epsilon &=&
   2\Gamma^\psi\partial_\psi\epsilon +\cos\theta\Gamma^\theta\epsilon ~.
 \end{eqnarray*}
 We can easily verify that (\ref{two local Lorentz})
 satisfies the above equations.
 The two unitary matrices of (\ref{two local Lorentz})
 are still the solution of (\ref{angular equation})
 if we replace $\Gamma^{\ul{1}}$ and $\Gamma^{\ul{3}}$
 with appropriate linear combinations of
 $\Gamma^{\ul{t,\sigma,1,2,3}}$,
 but the condition (\ref{residual SUSY}) fixes this arbitrariness.}
 The remaining equations (\ref{x_123 equation})
 determine the dependence on the coordinates $x_{1,2,3}$.
 Using (\ref{assumption for epsilon})
 and (\ref{residual SUSY}) we can rewrite them as follows:
\begin{equation}
 2\nabla_{\ul{i}}\epsilon = \nabla_{\ul{i}}\ln g_0\epsilon
\label{scale factor}
\end{equation}
\begin{equation}
  g_0(\frac{1}{2}\Omega_{\ul{ipq}}\gamma^{pq}
   -\nabla_{\ul{p}}\ln g_0\gamma^{ip})
 = E_i\gamma^{31}-B_i\gamma^{23}+H_i\gamma^{12}
\label{spin connection and field strength}
\end{equation}
\begin{equation}
  E_2=B_1=-H_3
\label{E2=B1=-H3}
\end{equation}
\begin{equation}
\begin{array}{rclclcl}
  3g_0\sla{\nabla}\ln g_0 &=&
   \gamma^1(-2    E_3 +~\;  H_2)  &\!\!+\!\!&
   \gamma^2(-     B_3 -~\,  H_1)  &\!\!+\!\!&
   \gamma^3(~~~\; B_2 +    2E_1) \\
  -3g_0g_\I^{-1}\gamma^1+3g_0\sla{\nabla}\ln g_\I &=&
   \gamma^1(~~~~  E_3 -    2H_2)  &\!\!+\!\!&
   \gamma^2(-     B_3 +    2H_1)  &\!\!+\!\!&
   \gamma^3(~~~\; B_2 -~\,  E_1) \\
  -3g_0g_\II^{-1}\gamma^3+3g_0\sla{\nabla}\ln g_\II &=&
   \gamma^1(~~~~  E_3 +~\;  H_2)  &\!\!+\!\!&
   \gamma^2(\;2   B_3 -~\,  H_1)  &\!\!+\!\!&
   \gamma^3(-2    B_2 -~\,  E_1)
\end{array}
\label{g_012}
\end{equation}
\comment{
 Here we have used the following equalities:
\[
  \gamma^{123}\epsilon=\gamma^{31}\epsilon~,~
  \gamma^0\epsilon = -\gamma^{23}\epsilon~,~
  \gamma^5\epsilon = -\gamma^{12}\epsilon.
\]
 and removed $\epsilon$ from the equations
 using the following lemma: \\
{\it  For real $a,b,c,d$ and nonzero $\epsilon$,~~}\\
  ~~~~~~~~~~
 $(a\gamma^1+b\gamma^2+c\gamma^3+d\gamma^{123})\epsilon=0
    ~~\Longrightarrow~~ a=b=c=d=0. $}
 The first equation (\ref{scale factor}) determines
 the scale factor of $\epsilon$ as follows:
\begin{equation}
  \epsilon= g_0^{1/2}U_\I U_\II\epsilon_0.
\end{equation}
 The next equations
 (\ref{spin connection and field strength}) and (\ref{E2=B1=-H3})
 relate the components of the spin connection
 and the gauge field strength:
\[
\begin{array}{rclcrclcrcl}
 \Omega_{\ul{112}}  &\!\!=\!\!& ~~g_0^{-1} H_1 + \nabla_{\ul{2}}\ln g_0 &;&
 \Omega_{\ul{212}}  &\!\!=\!\!& ~~g_0^{-1} H_2 - \nabla_{\ul{1}}\ln g_0 &;&
 \Omega_{\ul{312}}  &\!\!=\!\!& ~~g_0^{-1} H_3 \\
 \Omega_{\ul{123}}  &\!\!=\!\!&  -g_0^{-1} B_1                          &;&
 \Omega_{\ul{223}}  &\!\!=\!\!&  -g_0^{-1} B_2 + \nabla_{\ul{3}}\ln g_0 &;&
 \Omega_{\ul{323}}  &\!\!=\!\!&  -g_0^{-1} B_3 - \nabla_{\ul{2}}\ln g_0 \\
 \Omega_{\ul{131}}  &\!\!=\!\!& ~~g_0^{-1} E_1 - \nabla_{\ul{3}}\ln g_0 &;&
 \Omega_{\ul{231}}  &\!\!=\!\!& ~~g_0^{-1} E_2                          &;&
 \Omega_{\ul{331}}  &\!\!=\!\!& ~~g_0^{-1} E_3 + \nabla_{\ul{1}}\ln g_0 
\end{array}
\]
 Since the torsion-free condition (\ref{torsion-free condition})
 relates the spin connection to the vielbein,
 the above relations allow us to express
 the components of the gauge field strength
 in terms of the vielbein.
 The most elegant way is:
\begin{eqnarray}
  d\varpi^1 &=& H_1\varpi^{21}+E_1\varpi^{13} \nonumber \\
  d\varpi^2 &=& H_2\varpi^{21}+2E_2\varpi^{13}+B_2\varpi^{23} \nonumber \\
  d\varpi^3 &=& B_3\varpi^{23}+E_3\varpi^{13}
\label{elegant}
\end{eqnarray}
 The most convenient choice of coordinates
 that is compatible with the above would be
\begin{eqnarray}
  \varpi^1 &=& h_1dx^1 \nonumber \\
  \varpi^2 &=& h_2(dx^2+A_1dx^1+A_3dx^3) \nonumber \\
  \varpi^3 &=& h_3dx^3
\label{convenient choice}
\end{eqnarray}
\comment{
 In the dual representation
 (\ref{elegant}) and (\ref{convenient choice}) become
\[
\begin{array}{rcl}
  \left[\hat{\nabla}_3,\hat{\nabla}_1\right]
  &=&E_1\hat{\nabla}_1+2E_2\hat{\nabla}_2+E_3\hat{\nabla}_3 \\
  \left[\hat{\nabla}_3,\hat{\nabla}_2\right]
  &=&B_2\hat{\nabla}_2+B_3\hat{\nabla}_3 \\
  \left[\hat{\nabla}_1,\hat{\nabla}_2\right]
  &=&H_1\hat{\nabla}_1+H_2\hat{\nabla}_2
\end{array}
~~~~~
\begin{array}{rcl}
  \hat{\nabla}_1 &=& h_1^{-1}(\partial_1-A_2\partial_2) \\
  \hat{\nabla}_2 &=& h_2^{-1}\partial_2 \\
  \hat{\nabla}_3 &=& h_3^{-1}(\partial_3-A_3\partial_2)
\end{array}
\]
 where we have introduced $\hat{\nabla}_i=g_0\nabla_{\ul{i}}$
 for convenience.
 The components of the gauge field strength become
\begin{eqnarray*}
  E_1 &=& -\hat{\nabla}_3\ln h_1 \\
 2E_2 &=& h_2h_3^{-1}\hat{\nabla}_1A_3
                         -h_2h_1^{-1}\hat{\nabla}_3A_1 \\
  E_3 &=& \hat{\nabla}_1\ln h_3 \\
 2B_1 &=& h_2h_3^{-1}\hat{\nabla}_1A_3
                         -h_2h_1^{-1}\hat{\nabla}_3A_1 \\
  B_2 &=& h_2h_3^{-1}\hat{\nabla}_2A_3
                         -\hat{\nabla}_3\ln h_2 \\
  B_3 &=& \hat{\nabla}_2\ln h_3 \\
  H_1 &=& \hat{\nabla}_2\ln h_1 \\
  H_2 &=& h_2h_1^{-1}\hat{\nabla}_2A_1
                         -\hat{\nabla}_1\ln h_2 \\
-2H_3 &=& h_2h_3^{-1}\hat{\nabla}_1A_3
                             -h_2h_1^{-1}\hat{\nabla}_3A_1
\end{eqnarray*}
}
 The next equations (\ref{g_012}) enable us to express
 $g_{0,\I,\II}$ in terms of the components of $\varpi^i$.
 A careful analysis of them with the help of
 (\ref{elegant}) and (\ref{convenient choice})
 yields that $\frac{g_\I}{h_1g_0}$ depends only on $x_1$,
 and similarly $\frac{g_\II}{h_3g_0}$ depends only on $x_3$.
 Using the diffeomorphism degrees of freedom
 we can therefore set
\[
  g_\I=x_1h_1g_0 ~,~~~ g_\II=x_3h_3g_0.
\]
 Using the above relation we then find $g_0 g_\I g_\II=x_1x_3$.
 Hence
\begin{equation}
  g_0  = (h_1h_3)^{-\frac{1}{3}}~,~~~
  g_\I = x_1h_1^{\frac{2}{3}}h_3^{-\frac{1}{3}}~,~~~
  g_\II= x_3h_1^{-\frac{1}{3}}h_3^{\frac{2}{3}}
\end{equation}
 The equations (\ref{g_012}) also yield
 the following equations
\begin{equation}
 \partial_1 \left(\frac{h_2h_3}{h_1}\right)
 = \partial_2 \left(\frac{A_1h_2h_3}{h_1}\right)~,~~~~
 \partial_3 \left(\frac{h_2h_1}{h_3}\right)
 = \partial_2 \left(\frac{A_3h_2h_1}{h_3}\right)~.
\end{equation}
 Hence we put
\begin{equation}
  h_2^2 = \partial_2X\partial_2Y~,~~
  h_1^2 = \frac{\partial_2Y}{\partial_2Z}~,~~
  h_3^2 = \frac{\partial_2X}{\partial_2Z}~,~~
  A_1   = \frac{\partial_1X}{\partial_2X}~,~~
  A_3   = \frac{\partial_3Y}{\partial_2Y}.
\label{solution for BPS condition}
\end{equation}
 Thus the most generic solution of the BPS condition
 with spherical symmetry is summarized as follows:
\begin{eqnarray}
 ds^2 &=& \left(\partial_2X\partial_2Y\partial_2Z\right)^{-1/3}
   \left[\partial_2Z(-dt^2+d\sigma^2)
       + \partial_2Y(dx_1^2+x_1^2d\Omega_\I^2)
       + \partial_2X(dx_3^2+x_3^2d\Omega_\II^2)
         \right. \nonumber \\ && \left.~~~~~~~~~~~~~~~~~~~~~~~~~~
       +\partial_2X\partial_2Y\partial_2Z
    \left(dx_2+\frac{\partial_1X}{\partial_2X}dx_1
              +\frac{\partial_3Y}{\partial_2Y}dx_3\right)^2     \right] ~,
\label{spherical symmetric metric}
\end{eqnarray}
\begin{eqnarray}
  F_{(4)}&=&
   dtd\sigma\wedge \frac{1}{2}d\left[dx_1D_1Z+dx_3D_3Z\right]
      \nonumber \\ &&
  +d^3\Omega_\II\wedge \frac{1}{2}
   \left[-d(x_3^3D_3X)+x_3^3dx_3
          \left(\frac{\partial_2}{\partial_2Y}\frac{\partial_2X}{\partial_2Z}
                +x_3^{-3}D_3x_3^3D_3X\right)    \right]
      \nonumber \\ &&
  +d^3\Omega_\I\wedge \frac{1}{2}
   \left[-d(x_1^3D_1Y)+x_1^3dx_1
          \left(\frac{\partial_2}{\partial_2X}\frac{\partial_2Y}{\partial_2Z}
                +x_1^{-3}D_1x_1^3D_1Y\right)    \right] ~.
\label{spherical symmetric field strength}
\end{eqnarray}
 Here $D_1$ and $D_3$ are the ``coordinate covariant derivatives''
 defined as follows:
\[
 D_1 \equiv \partial_1-\frac{\partial_1X}{\partial_2X}\partial_2
     \equiv \partial_1|_{X {\rm fixed}}
  ~,~~~
 D_3 \equiv \partial_3-\frac{\partial_3Y}{\partial_2Y}\partial_2
     \equiv \partial_3|_{Y {\rm fixed}}~.
\]
 These are obviously invariant under the change of the coordinate
 $x_2 \rightarrow x_2' \!=\! f(x_1,x_2,x_3)$.
 This is a residual diffeomorphism symmetry,
 and owing to this symmetry we may parameterize
 the 2-direction by any of $X,Y,Z$.

\section{Generalization}
   From the previous result
 (\ref{spherical symmetric metric}), 
 (\ref{spherical symmetric field strength})
 we can guess the expression for more general solutions
 without spherical symmetry.
 It is expressed by three arbitrary functions
 $X, Y, Z$ of $x_{2,3,\ldots, \ten}$ as follows:
\begin{eqnarray}
 ds^2 &=& \left(\partial_2X\partial_2Y\partial_2Z\right)^{-1/3}
   \left[\partial_2Z(-dt^2+d\sigma^2)
       + \partial_2Y dx_i^2
       + \partial_2X dx_p^2
         \right. \nonumber \\ && \left.~~~~~~~~~~~~~~~~~~~~~~~~~~
       +\partial_2X\partial_2Y\partial_2Z
    \left(dx_2+\frac{\partial_iX}{\partial_2X}dx_i
              +\frac{\partial_pY}{\partial_2Y}dx_p\right)^2     \right]
\label{generic metric}
\end{eqnarray}
\begin{eqnarray}
  2F_{(4)}&=&
   dtd\sigma\wedge d\left[dx_iD_iZ+dx_pD_pZ\right]
      \nonumber \\ &&
  + \left[
    \frac{1}{6}\epsilon^{pqrs}d(D_pX) dx_qdx_rdx_s
    -dx_7dx_8dx_9dx_{10}\left(
       \frac{\partial_2}{\partial_2Y}\frac{\partial_2X}{\partial_2Z}
       +D_pD_pX\right)    \right]      \nonumber \\ &&
  + \left[
    \frac{1}{6}\epsilon^{ijkl}d(D_iY) dx_jdx_kdx_l
    -dx_3dx_4dx_5dx_6\left(
       \frac{\partial_2}{\partial_2X}\frac{\partial_2Y}{\partial_2Z}
       +D_iD_iY\right)    \right]
\label{generic field strength}
\end{eqnarray}
\[
 i,j,k,l= (3,4,5,6)~,~~~
 p,q,r,s= (7,8,9,10)~.
\]
 The coordinate covariant derivatives are defined as follows:
\[
 D_i
 \equiv \partial_i - \frac{\partial_iX}{\partial_2X}\partial_2
 \equiv \partial_i |_{X{\rm  fixed}} ~,~~~
 D_p \equiv \partial_p - \frac{\partial_pY}{\partial_2Y}\partial_2
 \equiv \partial_p |_{Y{\rm  fixed}} ~.
\]
 We can safely say that
 the above expression is the most generic BPS configuration,
 because it is the unique generalization of
 the most generic spherical symmetric configuration
 obtained in the previous section.
 We would like to note here again
 that one of $X,Y,Z$ is a residual diffeomorphism degree of freedom.

\section{Equation of Motion}
   The equation of motion in the absence of the source is given by
\begin{equation}
  dF_{(4)}=dF_{(7)}-F_{(4)}\wedge F_{(4)}=0
\label{Maxwell}
\end{equation}
\begin{equation}
  \frac{1}{4}\left[R_{mn}-\frac{1}{2}g_{mn}R\right] =
 \frac{1}{12}\left[F_{mpqr}F_n^{~~pqr}-\frac{1}{8}g_{mn}F_{pqrs}F^{pqrs}\right]
\label{Einstein}
\end{equation}
 A careful analysis of these equations shows that,
 under the assumption of the BPS condition
 some of the above equations turn out equivalent.
 The result is that the solution of (\ref{Maxwell})
 automatically satisfies (\ref{Einstein}).
 Therefore we concentrate on (\ref{Maxwell}) in the following.

   In the presence of the source the equation of motion is modified as
\begin{equation}
  dF_{(4)}=j_5~,~~~~
  dF_{(7)}-F_{(4)}\wedge F_{(4)}=j_8.
\end{equation}
 BPS condition now relates the components of the stress tensor
 to the components of $j_5$ and $j_8$.
 Our generic BPS configuration
 (\ref{generic metric}), (\ref{generic field strength})
 has the following currents:
\begin{eqnarray}
  2j_5 &=& -d^4x_{789\ten}\wedge df^{[M5]}
           -d^4x_{3456}   \wedge df^{[M5']}
 \label{j5} \\
  2j_8 &=& -d^8x_{3456789\ten} f^{[M2]}              \nonumber \\&&
    -d^4x_{789\ten}Dx_2 \partial_2Y\wedge
    \frac{1}{6}\epsilon^{ijkl} D_if^{[M5]}d^3x_{jkl} \nonumber  \\&&
    -d^4x_{3456}Dx_2    \partial_2X\wedge
    \frac{1}{6}\epsilon^{pqrs} D_pf^{[M5']}d^3x_{qrs} \nonumber  \\&&
    -dtd\sigma Dx_2d^4x_{3456}   \wedge\partial_2Z df^{[M5']}
    -dtd\sigma Dx_2d^4x_{789\ten}\wedge\partial_2Z df^{[M5]}
 \label{j8}
\end{eqnarray}
\[
  Dx_2 \equiv dx_2+ dx_i\frac{\partial_iX}{\partial_2X}
                  + dx_p\frac{\partial_pY}{\partial_2Y} ~,~~~
  d^nx_{j_1\cdots j_n} \equiv  dx_{j_1}\cdots dx_{j_n}~,
\]
 where the three functions $f^{[M2,M5,M5']}$ are defined as follows:
\begin{eqnarray}
 f^{[M5]} &=&
 \frac{\partial_2}{\partial_2Y}\frac{\partial_2X}{\partial_2Z} +D_pD_pX
    \nonumber \\
 f^{[M5']} &=&
  \frac{\partial_2}{\partial_2X}\frac{\partial_2Y}{\partial_2Z} +D_iD_iY \\
 f^{[M2]} &=&
      D_iD_i\frac{\partial_2X}{\partial_2Z}
     +D_pD_p\frac{\partial_2Y}{\partial_2Z}
     -2D_iD_pXD_pD_iY
     +\frac{\partial_2}{\partial_2Y}\frac{\partial_2X}{\partial_2Z}\cdot
      \frac{\partial_2}{\partial_2X}\frac{\partial_2Y}{\partial_2Z}
    \nonumber
\end{eqnarray}
 These encode the position of the sources.
 The source-free equations of motion are
 hence given by $f^{[M2]}\!=\!f^{[M5]}\!=\!f^{[M5']}\!=\!0$.

   If both M5-branes and M5'-branes are present,
 they possibly bend each other.
 However, bending of branes is a notion
 that depends on the choice of coordinates.
 We may say that there is no bending effects
 if we can find in a natural way a coordinate frame
 in which both M5 and M5'-branes are flat.
 But the following consideration leads us to conclude
 that this is not the case.

   Looking at the expressions for currents carefully,
 one finds that the fourth and the fifth terms
 in $j_8$ of (\ref{j8}) correspond to M2-brane charges
 with Euclidean worldvolume.
 Hence it is reasonable to require them to vanish
 even in the presence of the source.
 We therefore impose the following condition:
\begin{equation}
   D_if^{[M5]}\equiv D_pf^{[M5']}\equiv 0~.
\label{reasonable condition}
\end{equation}
 This is equivalent to requiring that
 $f^{[M5]}$  is a function of $(x_p, X)$ and
 $f^{[M5']}$ is a function of $(x_i, Y)$.
 Under the above condition the currents take
 the following simple form:
\begin{eqnarray*}
  2j_5 &=& -d^4x_{789\ten}Dx_2\partial_2f^{[M5]}
           -d^4x_{3456}   Dx_2\partial_2f^{[M5']} \\
  2j_8 &=& -d^8x_{3456789\ten} f^{[M2]} ~.
\end{eqnarray*}
 The classical solution for some isolated M5 and M5'-branes
 is thus obtained by solving
\begin{eqnarray}
 f^{[M5]}  =
 \frac{\partial_2}{\partial_2Y}\frac{\partial_2X}{\partial_2Z} +D_pD_pX
          &=& \sum_j Q_j \delta^4(x_p-a_p^{(j)})\theta(X-a_2^{(j)})
   \nonumber \\
 f^{[M5']} =
 \frac{\partial_2}{\partial_2X}\frac{\partial_2Y}{\partial_2Z} +D_iD_iY
          &=& \sum_j Q'_j\delta^4(x_i-b_i^{(j)})\theta(Y-b_2^{(j)}).
\label{M5 equation}
\end{eqnarray}
 The solution corresponds to the system of
 M5-branes of charge $Q_j$ at $(X,x_p)=(a_2^{(j)},a_p^{(j)})$
 and M5'-branes of charge $Q'_j$ at $(Y,x_i)=(b_2^{(j)},b_i^{(j)})$.
 We find that M5-branes are flat in $x_2=X$ frame
 while M5'-branes are flat in $x_2=Y$ frame.
 Hence we conclude that the M5-branes
 and M5'-branes in general bend each other.

   Choosing one of $X,Y,Z$ as the $x_2$-coordinate
 we can regard (\ref{M5 equation}) as two equations
 for two unknown functions.
   They are nonlinear and highly complicated equations,
 $(X,Y,Z)$ appearing as coordinates as well as functions.
 Moreover the solution of (\ref{M5 equation})
 must not be unique because there is a freedom
 to put an arbitrary number of M2-branes.
 At present the generic solution for them is not known.
 It is known, however, that under the assumption
\[
 \partial_2X        = H_5(x_p)~,~
 \partial_2Y        = H_{5'}(x_i)~,~
 (\partial_2Z)^{-1} = H_2(x_i,x_p).
\]
 the equations of motion are reduced
 to the following linear differential equations:
\[
 \partial_p\partial_p H_5    =
 \partial_i\partial_i H_{5'} =
 (H_5\partial_i\partial_i + H_{5'}\partial_p\partial_p)H_2 = 0 ~.
\]
 This type of equations has been analyzed
 in \cite{Itzhaki:1998uz, Surya:1998dx,
          Youm:1999zs, Marolf:1999uq} in different contexts.
 The above equations describe the system of M2-branes
 together with some M5 and M5'-branes
 smeared along the $x_2$-direction.
 Since all the fields are $x_2$-independent
 the solutions cannot represent M2-branes ending on M5-branes.

   The third equation $f^{[M2]}\!\!=\!\!0$ remains to be analyzed.
 In analyzing this, recall that one of $X,Y,Z$ is
 the gauge degree of freedom.
 Therefore if
 $f^{[M2]}\!\!=\!\!f^{[M5]}\!\!=\!\!f^{[M5']}\!\!=\!\!0$
 were three independent equations,
 the system would be over-determined.
 This is not the case.
 The point is that the $x_2$-derivative of $f^{[M2]}$
 is zero where $f^{[M5]}\!=\!f^{[M5']}\!=\!0$.
 Indeed, using (\ref{reasonable condition}) we find
\begin{equation}
  \partial_2f^{[M2]}=
  \partial_2f^{[M5]} \left(f^{[M5']}-D_iD_iY\right)
 +\partial_2f^{[M5']}\left(f^{[M5]} -D_pD_pX\right).
\label{M2 boundary}
\end{equation}
 Since $\partial_2f^{[M2]}$ represents
 the boundaries of M2-branes,
 the above equality means that M2-branes can have boundaries
 only on M5-branes.

\section{Brane Creation}
\begin{figure*}[htb]
\begin{center}
\epsfbox{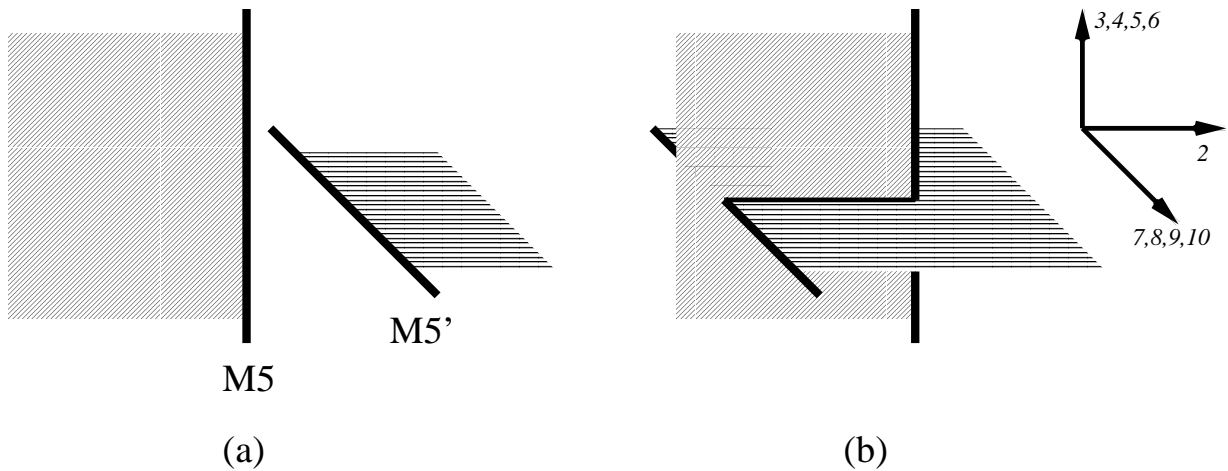}
\caption{Creation of an M2-brane by an M5 and an M5'-brane
         passing through each other.}
\label{branes}
\end{center}
\end{figure*}

   We would like to give an idea for
 how the brane creation can be seen in supergravity.
   Let us consider the system of an M5-brane and an M5'-brane.
 Then the functions $f^{[M5]}$ and $f^{[M5']}$
 have support on semi-infinite six-planes that are
 bounded by M5 and M5'-branes, respectively.
 Assume that one of the two six-planes is
 on the left of the M5-brane,
 and the other is on the right of the M5'-brane,
 as depicted in the Figure \ref{branes}.
 Note that one can change whether a six-plane
 appears on the left or on the right of an M5-brane
 by the shift $X \rightarrow X + f(x_p)$
 or $Y \rightarrow Y + f(x_i)$.
 Then, according to the relative position of
 two M5-branes the two six-planes may or may not
 have an intersection.
 Since the component $f^{[M2]}$ of
 the M2-brane current $j_8$ satisfies
\begin{eqnarray}
  \partial_2f^{[M2]}&=&
  \partial_2\left(f^{[M5]}f^{[M5']}\right)
  -\partial_2f^{[M5 ]}D_iD_iY
  -\partial_2f^{[M5']}D_pD_pX
  \label{M2 boundary 2}  \\
  &&\mbox{or}~~~~~
  f^{[M2]}= f^{[M5]}f^{[M5']}+\ldots, \nonumber
\end{eqnarray}
 there is an M2-brane precisely
 on the intersection of two six-planes,
 and its charge is proportional to the product
 of the charges of the two M5-branes.
 This explains the brane creation in supergravity,
 namely, when an M5-brane pass through an M5'-brane,
 an M2-brane is created between them.

   It is expected that all the other types of M2-branes,
 namely those with semi-infinite worldvolume
 or those stretching between M5-M5 or M5'-M5'
 are described by the second and third terms
 in (\ref{M2 boundary 2}).

   We give here a simple example.
 Let us solve the equation of motion
 under the assumption that $X,Y,Z$ depend only on $x_2$.
 The solution representing the system of
 an M5-brane at $Z\!=\!a$ and an M5'-brane at $Z\!=\!b$
 is obtained by solving
\begin{equation}
  \frac{\partial_Z^2X}{\partial_ZY} = \theta(Z-a)  ~,~~~
  \frac{\partial_Z^2Y}{\partial_ZX} =-\theta(b-Z)  ~.
\label{x_2 equations}
\end{equation}
 Assuming $a<b$, the solution is given
 in terms of a function $f(Z)$ satisfying $f''(Z)=-f(Z)$
 as follows:
\[
\begin{array}{l}
 (Z\le a) \\
 \left\{\begin{array}{l}
 \partial_ZX = f(a) \\
 \partial_ZY = f''(a)(Z-a)+f'(a)
 \end{array}\right.
\end{array}
~~
\begin{array}{l}
 (a\le Z \le b)  \\
 \left\{\begin{array}{l}
 \partial_ZX = f(Z) \\
 \partial_ZY = f'(Z)
 \end{array}\right.
\end{array}
~~
\begin{array}{l}
 (b\le Z)  \\
 \left\{\begin{array}{l}
 \partial_ZX = f'(b)(Z-b)+f(b) \\
 \partial_ZY = f'(b)
 \end{array}\right.
\end{array}
\]
 Then $f^{[M2]}$ takes the following form as expected:
\begin{equation}
  f^{[M2]}=
 \frac{\partial_Z^2X}{\partial_ZY}
 \frac{\partial_Z^2Y}{\partial_ZX}=-\theta(Z-a)\theta(b-Z)~.
\end{equation}
 This represents the M2-brane stretching
 between the two M5-branes, completely de-localized
 in the $x_{3,4,\ldots,9,\ten}$-directions.
 If the right-hand sides of the equations
 (\ref{x_2 equations}) are shifted by constants,
 the solutions will contain some M2-branes
 with semi-infinite worldvolume.
 It is straightforward to find such solutions.

\section{Conclusion}
   In this article we have found
 the most generic BPS configuration
 for M5-branes(013456), M5'-branes(01789$\ten$)
 and M2-branes(012).
 We have also given and studied
 the equation of motion for localized sources.
 The equations are highly nonlinear, and it seems
 very difficult to obtain the generic solution.
 In fact, it is not clear whether or not the solution
 for localized M5 and M5'-branes indeed exists.
 But the analysis of the equations of motion themselves
 has lead to some interesting results.

   By focusing on a specific term in the M2-brane current
 we have given an explanation for the brane creation
 in supergravity.
 Strictly speaking, however, this is no more than
 a conjecture because we have no justification for
 picking up a specific term in the current.
 Constructing a solution for the equations (\ref{M5 equation})
 will help us in great deal in understanding
 the mechanism of brane creation in supergravity
 and checking if the above conjecture indeed holds.

~

\noindent{\large\bf Acknowledgment}\\
   The author thanks J. Hashiba for collaboration
 at the early stage of this work.
 The author is also thankful to
 T.~Eguchi, Y.~Sugawara and S.~Terashima
 for discussions and comments.
 The work of the author was supported in part by
 JSPS Research Fellowships for Young Scientists.

\end{document}